\documentclass[sigconf]{acmart}

\usepackage{amssymb,amsfonts}
\usepackage{algorithmic}
\usepackage{textcomp}
\usepackage{multirow}
\usepackage{url}
\usepackage{enumitem}
\usepackage{siunitx}
\usepackage{ifthen}
\usepackage{soul}
\usepackage{subcaption}
\usepackage{makecell}
\usepackage{indentfirst}

\newlength{\taxw}
\makeatletter
\newcommand{\midfootnotesize}{\@setfontsize\midfootnotesize{7.5pt}{9pt}}
\makeatother

\definecolor{mygrayzero}{gray}{0.9}
\definecolor{mygrayone}{gray}{0.8}
\definecolor{mygraytwo}{gray}{0.7}
\definecolor{mygraythree}{gray}{0.6}

\def\BibTeX{{\rm B\kern-.05em{\sc i\kern-.025em b}\kern-.08em
    T\kern-.1667em\lower.7ex\hbox{E}\kern-.125emX}}

\copyrightyear{2026}
\acmYear{2026}
\setcopyright{acmlicensed}
\acmConference[ICPC '26]{34th IEEE/ACM International Conference on Program Comprehension}{2026}{Brazil}
\acmBooktitle{34th IEEE/ACM International Conference on Program Comprehension (ICPC '26), Brazil}
\acmDOI{XX.XXXX/XXXXXXX.XXXXXXX}
\acmISBN{XXX-X-XXXX-XXXX-X/XXXX/XX}

\begin{document}

\title{Algorithm-Based Pipeline for Reliable and Intent-Preserving Code Translation with LLMs}

\author{Shahriar Rumi Dipto}
\affiliation{
   \institution{University of Saskatchewan, Canada}
   \city{}
  \country{}
}
\email{rumi.dipto@usask.ca}

\author{Saikat Mondal}
\affiliation{
   \institution{University of Saskatchewan, Canada}
   \city{}
  \country{}
}
\email{saikat.mondal@usask.ca}

\author{Chanchal K. Roy}
\affiliation{
   \institution{University of Saskatchewan, Canada}
   \city{}
  \country{}
}
\email{chanchal.roy@usask.ca}

\renewcommand{\shortauthors}{Dipto et al.}

\begin{abstract}
Code translation, the automatic conversion of programs between languages, is a growing use case for Large Language Models (LLMs). However, direct one-shot translation often fails to preserve program intent, leading to errors in control flow, type handling, and I/O behavior.  
We propose an algorithm-based pipeline that introduces a language-neutral intermediate specification to capture these details before code generation.
This study empirically evaluates the extent to which structured planning can improve translation accuracy and reliability relative to direct translation. 
We conduct an automated paired experiment -- \textit{direct} and \textit{algorithm-based} to translate between Python and Java using five widely used LLMs on the Avatar and CodeNet datasets. For each combination (model, dataset, approach, and direction), we compile and execute the translated program and run the tests provided. We record compilation results, runtime behavior, timeouts (e.g., infinite loop), and test outcomes. We compute accuracy from these tests, counting a translation as correct only if it compiles, runs without exceptions or timeouts, and passes all tests. We then map every failed compile-time and runtime case to a unified, language-aware taxonomy and compare subtype frequencies between the direct and algorithm-based approaches.
Overall, the Algorithm-based approach increases micro-average accuracy from 67.7\% to 78.5\% ($\uparrow$10.8\%). It eliminates lexical and token errors by 100\%, reduces incomplete constructs by 72.7\%, and structural and declaration issues by 61.1\%. It also substantially lowers runtime dependency and entry-point failures by 78.4\%.
These results demonstrate that algorithm-based pipelines enable more reliable, intent-preserving code translation, providing a foundation for robust multilingual programming assistants.

\end{abstract}



\maketitle

\section{Introduction} 
\label{Introduction}

Reliable and intent-preserving code translation across programming languages has become a critical challenge in modern software engineering (SE) \cite{yang2024exploring}. Developers increasingly rely on automated translation to reuse code, migrate legacy systems, and develop applications across platforms \cite{Weisz2021PerfectionNRA,Weisz2022BetterTAA,Macedo2024ExploringTIA}. However, translated programs often contain semantic bugs, compile-time/runtime failures, or logic drift, making them unreliable \cite{yin2024rectifier,pan2024lost}. Large-scale studies report that up to 77.8\% of translated programs do not compile or produce incorrect results \cite{pan2024lost}. These failures commonly arise from improper handling of data types, control flows, boundary conditions, or I/O operations, which distort the intended algorithmic behavior of the source program \cite{Jana2023CoTranALA,yin2024rectifier}. 
Such faults increase debugging costs, introduce hidden defects, and reduce developer confidence in automated translation pipelines \cite{roziere2020unsupervised}. As translation systems become integral to software modernization \cite{bearblogSingleChatGPT}, it is essential to mitigate these reliability gaps. 
This work advances a reliability-oriented approach to LLM translation by introducing an algorithmic pipeline that preserves intent and reduces systematic failure modes.

Large Language Models (LLMs) have made code translation more practical and efficient by learning from large and diverse code corpora.
Despite strong fluency and generalization, recent studies \cite{pan2024lost,Farrukh2025SafeTransLTA,yin2024rectifier} show that LLM-translated code still suffers from frequent compilation errors and functional inconsistencies, with overall accuracy ranging only from 2.1\% to 47.3\% \cite{Yang2024ExploringAU,pan2024lost}.

The model often preserves the high-level algorithmic structure, but fails to capture critical low-level details. Slight deviations in type handling or I/O semantics can break an otherwise correct algorithm. For example, when translating between languages with different numeric parsing or division rules, an LLM may retain the loop structure but alter boundary checks or off-by-one conditions. Similarly, input patterns that assume a particular tokenization or newline behavior can reorder operations or modify array dimensions, producing code that compiles but violates the intended logic of the source program \cite{yin2024rectifier,Jana2023CoTranALA}. 
A task that should be a simple code transfer often turns into long rounds of testing and fixing. Without a way to keep the original algorithm’s purpose clear during translation, teams end up with fragile code and lose trust in the process \cite{Zhang2024ScalableVCA}.
To address this, a translation framework that first captures a concise, language-neutral algorithm, including its inputs, data structures, numeric rules, loop bounds, and output format, can help preserve intent, reduce failures, and restore reliability at scale.

Prior approaches to automated code translation fall into three main categories, but none explicitly enforce algorithmic intent before generation. First, rule-based and analysis-guided transpilers \cite{gotranspile_cxgo,immunant_c2rust,paulirwin_JavaToCSharp_2025,mono_sharpen} encode explicit language-pair rules, but they often produce non-idiomatic or unsafe code that requires extensive manual clean-up \cite{Ibrahimzada2025, liu2025llmigrate}. This rigidity and upkeep burden make pure rule-based approaches ill-suited for evolving languages and coding practices. Second, neural and unsupervised models trained on large code corpora \cite{roziere2020unsupervised,roziere2021leveraging,szafraniec2022code} outperformed traditional transpilers on benchmark tasks but still lack subtle semantic alignment, often missing corner-case behaviors or misusing operators and data types. Such errors indicate that purely neural approaches may generate functionally incorrect code if the source’s intent isn’t fully captured \cite{roziere2021leveraging}. Third, LLM-centric pipelines perform post-hoc repair using test or compiler feedback, as in UniTrans \cite{yang2024exploring,fan2023large}, or adopt bridge-language strategies inspired by natural language translation, such as InterTrans \cite{Kim2019PivotbasedTL,wu-wang-2007-pivot,fan2021beyond,macedo2025intertrans}. However, these methods generally translate first and repair or route later, without explicitly enforcing a compact, language-agnostic algorithm at inference time. In other words, current LLM pipelines excel at post-hoc error correction but do not inherently guarantee that the first-pass translation follows all the source’s semantic contracts.

In natural language translation, difficult language pairs are often handled through a pivot language (e.g., French$\rightarrow$English$\rightarrow$German) \cite{wu-wang-2007-pivot}. This remains effective in multilingual machine translation, where bridge languages help scale across language families without exhaustive pairwise training \cite{fan2021beyond}. Motivated by this principle, our work explicitly forces a language-agnostic algorithmic blueprint at inference time before final code generation. No prior work has taken this inference-stage approach as a proactive safeguard for semantic fidelity. Earlier systems either translate via another language (bridging through actual code in a pivot PL) \cite{macedo2025intertrans} or apply repairs after translation \cite{yin2024rectifier}. By locking down the correctness of the algorithm upfront, our approach can then generate idiomatic code in the target language that already adheres to those invariants.

To achieve this goal, we compared two LLM-based translation workflows: a direct pipeline (i.e., approach) and an algorithm-based pipeline that first generates a language-agnostic algorithm capturing I/O contracts, data structures, numeric rules, loop bounds, and output format before producing the target code. We evaluated both translation directions using the Avatar and CodeNet datasets with five LLMs: DeepSeek R1, DeepSeek V3, Llama 4 Maverick, GPT-4o, and Qwen2.5. For each combination of model, dataset, approach, and direction (e.g. DeepSeek-R1 vs. Avatar vs. Direct vs. Python$\rightarrow$Java), we compiled the translated programs, executed the provided tests, and recorded three outcomes: compile success/failure, runtime success/failure, and test success/failure. Failures were categorized into four groups: compile errors, runtime errors, test-only mismatches, and timeouts (e.g., infinite loop). Each unsuccessful case was then mapped to a unified, language-aware error taxonomy applied consistently across all models, datasets, and directions. Finally, we aggregated subtype counts for both the Direct and Algorithm approaches, computed frequencies and percentages, and marked a subtype as mitigated when the Algorithm error rate was lower than the Direct error rate. This taxonomy-driven comparison reveals where the algorithm reduces specific failure modes and where challenges persist. 
In particular, we answered three research questions and made three contributions.

\noindent\textbf{RQ\textsubscript{1}) To what extent does an algorithm-based pipeline improve code translation performance relative to direct one-shot translation?} 
This research question examines whether adding an algorithmic step improves reliability when translations must preserve control flow, type semantics, and I/O contracts. We conducted paired Direct and Algorithm-based translations on Avatar and CodeNet datasets in both Python$\rightarrow$Java and Java$\rightarrow$Python, recording compile success, runtime stability, and test accuracy. The algorithm-based approach achieved more consistent results across models, datasets, and directions, with clearer gains where structural precision matters, improving micro-average accuracy from 67.7\% to 78.5\% ($\uparrow$10.8\%).

\noindent\textbf{RQ\textsubscript{2}) What is the taxonomy and frequency distributions of compile-time and runtime error types observed in each translation pipeline?}
This research question goes beyond overall accuracy to examine why translations fail, focusing on tokenization, typing, indexing, imports, and program start-up or termination. We mapped each unsuccessful compile-time and runtime case to a unified, language-aware taxonomy applied consistently across models, datasets, and directions, and computed their frequencies. Parsing and Type Conversion errors dominate runtime failures (55\%) and remain the top category for both Direct and Algorithm-based pipelines. At compile-time, Structural and Declaration Issues are most frequent (54.6\%), particularly in Python$\rightarrow$Java, showing that the main challenge is preserving Java-style scaffolding and highlighting the need for tighter file, class, and method templates guided by grammar-constrained generation.

\noindent\textbf{RQ\textsubscript{3}) In which scenarios or code patterns does the algorithm-based pipeline reduce or eliminate specific error types compared to direct translation?} 
This research question examines whether a language-neutral algorithm sketch before code generation reduces the error modes that break fidelity in Python$\rightarrow$Java and Java$\rightarrow$Python translation. Using the same paired setup as RQ2 on Avatar and CodeNet, we aggregated taxonomy counts for Direct and Algorithm pipelines and compared their frequencies. The algorithm-based approach shows clear reductions: runtime dependency and entry-point errors drop from 89.2\% to 10.8\%, while compile-time lexical and token errors fall from 100\% to 0\%, incomplete constructs from 86.4\% to 13.6\%, and structural and declaration issues from 80.6\% to 19.4\%. A smaller group of errors, such as type or overload resolution and missing state or invalid references, persist or worsen, showing where the algorithm helps less and where tighter semantic constraints remain needed.

\smallskip
\noindent We position the algorithm-based pipeline as a general paradigm for reliable and intent-preserving code translation, shifting LLMs from token-level generation to contract-driven realization.

\smallskip
\noindent \textbf{Replication Package} can be found in our online appendix \cite{replication-package}.

\begin{figure}[h]
  \centering
  \includegraphics[width=3.4in]{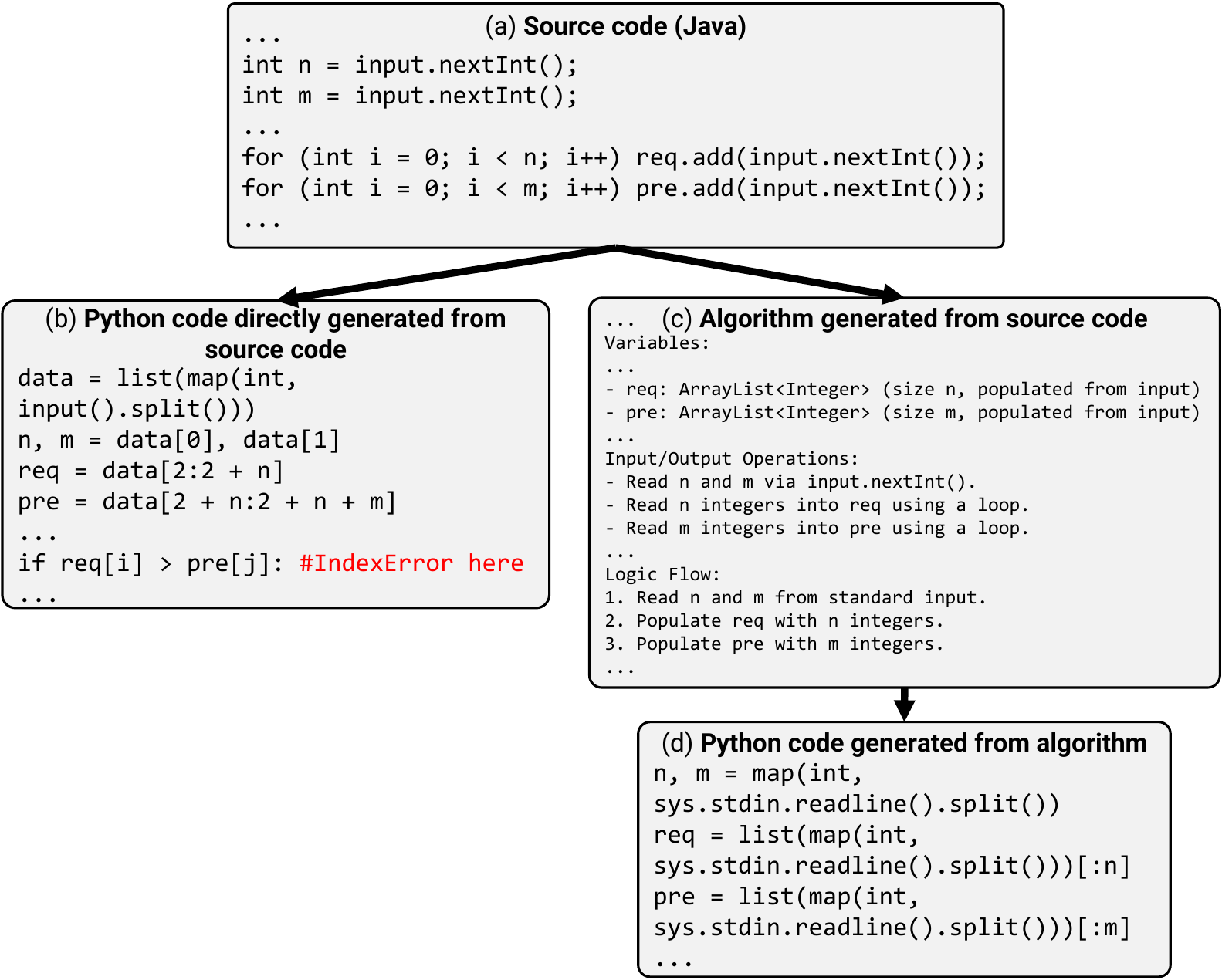}
  \caption{Motivating example showing how algorithm-based translation reduces errors compared to direct translation.}
  \label{fig:compare_src_translated_code}
  \vspace{-4mm}
\end{figure}

\section{Motivating Example}
This section provides an example of how algorithm-based translation prevents a common I/O pitfall in Java$\rightarrow$Python translation.

The Java program (see Figure \ref{fig:compare_src_translated_code}a) first reads two numbers, then reads two lists, and finally it scans those lists from the end to compute the answer.
A direct Python translation breaks this contract (see Figure \ref{fig:compare_src_translated_code}b). It assumes all input comes on one line, builds lists that are too short, and then Python raises \textit{IndexError} (out-of-range) when it tries to access items that are not there.
On the other hand, the algorithm-based approach avoids this mistake (see Figure \ref{fig:compare_src_translated_code}d). The algorithm (see Figure \ref{fig:compare_src_translated_code}c) solved a problem of input clarity. It explicitly stated what to read, in what order, and how many values, so the code collects the exact data before any processing. Reading the counts first and then gathering exactly that many items keeps list sizes correct, prevents out-of-bounds lookup, and makes the downstream logic run reliably.
This is a representative example of many. Therefore, a comprehensive study is required to explore how algorithmic representations can overcome these challenges and enable more accurate and reliable code translation.

\section{Methodology}

\begin{figure*}[h]
  \centering
  \includegraphics[width=6in]{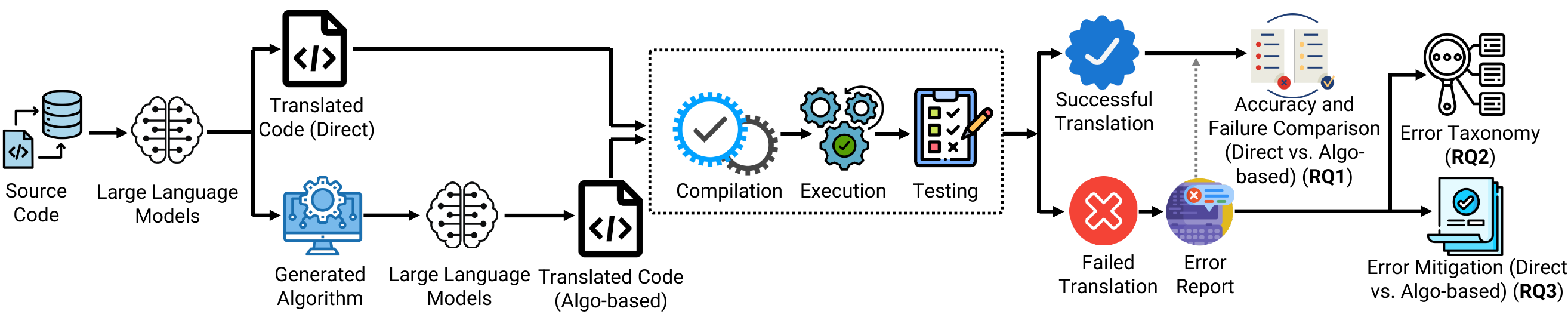}
  \caption{Study Methodology}
  \label{fig:methodology}
  \vspace{-3mm}
\end{figure*}

Figure \ref{fig:methodology} shows the overall methodology of our exploratory study. The details of each step are discussed below.

\subsection{Dataset Preparation}

We evaluated on pre-processed subsets of \textit{Avatar} \cite{ahmad2021avatar} and \textit{CodeNet} \cite{puri2021codenet} curated by Pan et al. \cite{pan2024lost}, widely used in recent LLM translation work \cite{nitin2024spectra, yin2024rectifier, zhu2024semi}. These splits provided validated tests and clean input–output pairs, allowing fully automated compilation and execution for a fair, like-for-like comparison between pipelines. Avatar is a parallel corpus of aligned Java-Python solutions for competitive programming, ideal for a translation task between a statically and a dynamically typed programming language. CodeNet is a large multi-language corpus. We selected the Python and Java subsets, with each program linked to a problem description and paired with functional tests. This gives a broad coverage of algorithmic tasks and enables automatic correctness checks.

We focused on Python and Java in both datasets for our study of cross-language code translation. These languages rank among the top four worldwide in the TIOBE index \cite{tiobe2024}. They also reflect different programming paradigms: Python is dynamically typed and interpreted, while Java is statically typed and compiled. This difference makes translation non-trivial \cite{jiao2023evaluationofnmt}, requiring models to bridge structural, syntactic, and semantic differences. By evaluating both directions on Avatar and CodeNet with the same automated evaluation pipeline for compilation, execution, and testing, we assessed how well LLMs handle cross-paradigm translation and structural mismatches between languages.

\subsection{LLM Models Used in Code Translation}
We evaluated five top general-purpose LLMs from OpenRouter \cite{openrouter}, chosen to cover diverse architectures. Selection was guided by the BigCodeBench leaderboard (May 2025) \cite{bigcodebenchleaderboard} and OpenRouter rankings (May 2025) \cite{openrouterRankings}, where these models appear near the top, indicating strong capabilities. Moreover, these LLMs are widely used in many software engineering tasks such as code generation and code translation, where they performed promisingly \cite{xu2025embedagent, xue2025classeval, qing2025effibench}. 

$\blacktriangleright$ \textit{DeepSeek R1} \cite{guo2025deepseek} is a reasoning-oriented model trained with supervised fine-tuning and multi-stage RL. We use it as a strong general baseline for structured problem solving.
$\blacktriangleright$ \textit{DeepSeek V3} \cite{liu2024deepseek} is a large Mixture-of-Experts model with state-of-the-art open-source results on code and math, providing a top-tier reference. 
$\blacktriangleright$ \textit{Llama 4 Maverick} \cite{metaLlamaHerd} is a widely available open model not specifically tuned for code, included to gauge general-purpose open-source performance.
$\blacktriangleright$ \textit{GPT-4o} \cite{hurst2024gpt} is a proprietary model with strong zero-shot performance in software engineering, serving as a state-of-the-art commercial reference. 
$\blacktriangleright$ \textit{Qwen2.5 (72B-Instruct)} \cite{qwen2025qwen25technicalreport} is a high-capacity dense model with strong code and math skills and long-context support, serving as a competitive high-end baseline.

\subsection{Translation Pipelines}
To investigate the behavior of LLMs in automated code translation, we defined two parallel translation workflows: (1) Direct code translation and (2) Algorithm-based translation using a language-agnostic intermediate representation. We studied these pipelines because direct translation is common in practice, yet it often drifts on I/O operations, types, bounds, and output. While on the other hand, the algorithm-based pipeline adds a brief language-neutral algorithm to lock these details before coding, so we can test if it preserves intent and reduces compile or runtime failures. 
Both workflows were applied to source code samples from the Avatar and CodeNet datasets and operated bidirectionally (Python$\rightarrow$Java and Java$\rightarrow$Python). Each method was independently evaluated using identical test cases and performance metrics, enabling a controlled comparison between translation strategies. 

We used step-by-step, chain-of-thought prompting \cite{wei2022chain} to make the model reason through the task rather than copy surface code patterns. It is well-suited to code translation in our setting because most failures come from hidden constraints. It makes the model spell out these constraints before it writes code, which directly targets our top error sources \cite{Wang2024DoALA}. In the Direct pipeline, the prompted lists provide brief analysis steps but still ask the model to produce the target code in one pass. In the Algorithm-based pipeline, we first asked for a language-agnostic algorithm, then used a second prompt to turn that algorithm into code in the target language. Prompts were fixed across models and datasets. We evaluated only the final code (compile, runtime, tests), and the intermediate chain-of-thought text was stored for analysis but not graded.

\begin{table}[!htbp]
\caption{Prompt template used in the direct code translation}
\label{tab:direct_translation_prompt_template}
\centering
\resizebox{3.15in}{!}{%
    \begin{tabular}{p{10cm}}
    \toprule
    \textcolor{blue}{<SOURCE\_LANG>} code: \textcolor{red}{<SOURCE\_CODE>}\\
    \\
    
        Let's think step-by-step and translate the above \textcolor{blue}{<SOURCE\_LANG>} code into functionally equivalent \textcolor{blue}{<TARGET\_LANG>} code using the following instructions: \\
        \textbf{Step 1:} Analyze the structure and identify main components, control flow, data types, and dependencies \\
        \textbf{Step 2:} Map \textcolor{blue}{<SOURCE\_LANG>} constructs to \textcolor{blue}{<TARGET\_LANG>} equivalents and determine required libraries \\
        \textbf{Step 3:} Generate \textcolor{blue}{<TARGET\_LANG>} code preserving exact functionality, names, logic, and program structure \\
        \textbf{Step 4:} Finally, provide only the \textcolor{blue}{<TARGET\_LANG>} code without any headers, comments, explanations, or examples \\
    \bottomrule
    \end{tabular}
    }
    \vspace{-2mm}
\end{table}

\subsubsection{\textbf{Direct Pipeline}}
In the direct translation approach, the source program (Python or Java) and a prompt (Table \ref{tab:direct_translation_prompt_template}) were provided to the LLM, which generated the target-language code in a single step. This mirrors common practice, treating the LLM as a bilingual code generator that can reuse surface cues from the source (e.g., variable names, data structures, control flow). This workflow serves as the baseline for assessing one-shot source-to-target language mapping without decomposition.

\begin{table}[!htbp]
\caption{Prompt template used for algorithm generation}
\label{tab:algo_generation_prompt_template}
\centering
\resizebox{3.15in}{!}{%
    \begin{tabular}{p{10cm}}
    \toprule
        \textcolor{blue}{<SOURCE\_LANG>} code: \textcolor{red}{<SOURCE\_CODE>}\\
        \\
        Let's think step-by-step and extract a detailed algorithm from the above \textcolor{blue}{<SOURCE\_LANG>} code using the following instructions: \\
        \textbf{Step 1:} Identify the code structure and analyze function signatures, classes, and program organization \\
        \textbf{Step 2:} Extract the core logic and trace execution flow, control structures, data transformations, and I/O operations \\
        \textbf{Step 3:} Document the algorithm components and specify data types, dependencies, and exact conditions \\
        \textbf{Step 4:} Finally, provide only the detailed algorithm without any headers, comments, explanations, or examples \\
    \bottomrule
    \end{tabular}
    }
    \vspace{-2mm}
\end{table}

\begin{table}[!htbp]
\caption{Prompt template used to generate code from algorithm}
\label{tab:code_generate_from_algo_prompt_template}
\centering
\resizebox{3.15in}{!}{%
    \begin{tabular}{p{10cm}}
    \toprule
        \textcolor{blue}{algorithm:} \textcolor{red}{<ALGORITHM>} \\
        \\
        Let's think step-by-step and generate complete, executable \textcolor{blue}{<TARGET\_LANG>} code from the above algorusing the following instructions: \\
        \textbf{Step 1:} Analyze the algorithm requirements and identify data structures, input/output specifications, and core logic components \\
        \textbf{Step 2:} Design the \textcolor{blue}{<TARGET\_LANG>} implementation and plan program structure, determine classes/functions, select appropriate data types, and language constructs \\
        \textbf{Step 3:} Generate the complete code and implement logic flow with proper syntax, add necessary imports/dependencies, handle edge cases, and ensure code compiles and executes correctly \\
        \textbf{Step 4:} Finally, provide only the \textcolor{blue}{<TARGET\_LANG>} code without any headers, comments, explanations, or examples \\
    \bottomrule
    \end{tabular}
    }
    \vspace{-2mm}
\end{table}

\subsubsection{\textbf{Algorithm-Based Pipeline}}
\label{section:algo_based_pipeline}
This approach separated reasoning from code generation to test if an explicit algorithm step improves reliability and correctness. The workflow has two stages:
\begin{itemize}
	\item \textbf{Algorithm extraction.} We designed a structured prompt (Table \ref{tab:algo_generation_prompt_template}) through several iterations to guide the LLM in producing a step-by-step, language-agnostic algorithm that captures the program’s logic, including control flow, data operations, and I/O handling. This design reduced ambiguity and focused the model on understanding and implementing the intended logic. It tested the model’s ability to generate code from logical specifications rather than translating code line by line.

	\item \textbf{Target code generation.} We then fed that algorithm back to the LLM with a second prompt (Table \ref{tab:code_generate_from_algo_prompt_template}) to produce code in the target language. The model was instructed to implement the specification using idiomatic syntax and types, rather than mapping lines directly.

\end{itemize}

\subsection{\textbf{Evaluation}}
After translation, we archived all artifacts for evaluation. For each model on Avatar and CodeNet, we used the same automated process to compile the generated code, execute it on the dataset's test suite, and record structured logs. From these logs, we produced summaries with accuracy, counts of compile failures, runtime failures, test case failures, and timeouts. We also kept error traces so failed runs can be mapped into our runtime and compile-time taxonomies. These standardized reports drive comparisons across models, datasets, directions, and pipelines. The process is fully automated and performed independently for every combination defined by model, dataset, approach (Direct or Algorithm-based), and translation direction (Python$\rightarrow$Java or Java$\rightarrow$Python). A translation was considered correct only when it compiled successfully, executed without exceptions or timeouts, and passed all test cases. These results provided consistent measures of compilation success, runtime stability, and functional correctness across combinations.

\subsubsection{\textbf{Performance Comparison: Direct Vs. Algorithm-Based Translation (RQ1)}}
For each combination, we reported the total number of instances, the number of correct translations, and the failures divided into four broad categories: (1) \textit{compile-time error} (build fail due to syntax/imports/type issues), (2) \textit{runtime error} (program crashes or raises an exception while running), (3) \textit{test-case failure} (program runs but produces incorrect outputs on tests), and (4) \textit{timeout} (program does not terminate within the time limit). This provided the top-level accuracy and failure profile used for cross-combination comparisons. 

In addition to per-combination reporting, we computed a micro-averaged translation accuracy by pooling all instances across models, datasets, directions, and approaches to summarize overall performance across the entire evaluation set. Micro-averaging is widely used to produce a single headline score over heterogeneous groups \cite{Turner2013AutomatedAO,DeDiego2022GeneralPS}. We adopted the same pooling idea, but applied it to correct vs. incorrect outcomes in code translation rather than per-label F1.

Let \(\mathcal{C}\) be the set of all combinations (model, dataset, approach, and direction), and for each combination \(c \in \mathcal{C}\) let \(N_c\) denote the number of instances and \(K_c\) the number of correct translations. For any subset of combinations \(S \subseteq \mathcal{C}\) (e.g., all Direct runs or all Algorithm runs), the micro-average accuracy is

\begin{equation}
\mathrm{Acc}_{\mathrm{micro}}(S) \;=\; 
\frac{\sum_{c \in S} K_c}{\sum_{c \in S} N_c}.
\label{eq:micro_accuracy}
\end{equation}
We used this pooled measure for two reasons: (i) it reflects the probability that a randomly drawn instance from the full workload is translated correctly (deployment-oriented view), and (ii) it respects size imbalance across combinations (large cells contribute proportionally more evidence than small cells), avoiding the equal-weighting bias of macro-averages.

\subsubsection{\textbf{Error Taxonomy and Frequency Analysis (RQ2)}}
All unsuccessful translations in the compile-time and runtime buckets were mapped to a consistent taxonomy of subtypes appropriate to the target language, and the same definitions were applied uniformly across models, datasets, and directions. To build this taxonomy in a reproducible way, we used well-established qualitative methods. We discovered and refined categories with thematic analysis and the constant comparative method by coding failures, comparing similar cases, and merging or splitting categories as needed \cite{braun2006using, corbin2014basics}. To keep the process rigorous for software engineering data, we followed codebook-based content analysis guidelines by writing a clear codebook with inclusion and exclusion rules, short definitions, and direction-aware examples for each runtime and compile-time category, following Cruzes et al.'s guideline \cite{cruzes2011recommended}. Two annotators led this process. They first drafted the codebook, then ran a small pilot in which both independently labeled the same failures ("double-coding") to test clarity and consistency. They discussed and resolved disagreements and then froze the taxonomy before labeling the entire set, following standard practice for category stability \cite{krippendorff2018content}. After the pilot, the annotators refined the codebook, re-coded the pilot items, and then labeled the full dataset independently. Each failed translation receives one label at the earliest failure stage. We report inter-rater reliability using Cohen's kappa and interpret it with accepted thresholds \cite{cohen1960coefficient, byrt1996good}. In our study, the double-coded subset reached substantial agreement (Cohen's kappa = 89.3\%), exceeding the commonly accepted 70\% threshold \cite{byrt1996good}, and the remaining disagreements were resolved through discussion and agreement. Finally, we applied a simple precedence rule (compile-time before runtime) that fits our scenario, avoiding double-counting and keeping labeling consistent in both Python$\rightarrow$Java and Java$\rightarrow$Python directions.

For each approach (Direct and Algorithm), we counted the number of failures in each compile-time and runtime subtype and reported both frequencies and percentages. For frequency, we computed the grand-level proportions by summing raw counts across all combinations and then normalizing by the corresponding grand totals. Let $\mathcal{C}$ be the set of all combinations (model, dataset, approach, and direction). For each runtime subtype $\mathrm{RE}_k \in \{\mathrm{RE}_1,\dots,\mathrm{RE}_6\}$, let $n_{\mathrm{RE}_k}(c)$ denote its count in combination $c$, treating cells marked “$\times$” as zero. The grand total for runtime and the overall frequency of subtype $\mathrm{RE}_k$ are
\begin{equation}
N_{\mathrm{RE}} = \sum_{c \in \mathcal{C}} \sum_{k=1}^{6} n_{\mathrm{RE}_k}(c),
\qquad
f_{\mathrm{RE}_k} = \frac{\sum_{c \in \mathcal{C}} n_{\mathrm{RE}_k}(c)}{N_{\mathrm{RE}}}.
\label{eq:overall_runtime_error_freq}
\end{equation}
We applied the same procedure at compile-time. For each compile subtype $\mathrm{CE}_j \in \{\mathrm{CE}_1,\dots,\mathrm{CE}_7\}$ with counts $n_{\mathrm{CE}_j}(c)$, the grand total and frequencies are
\begin{equation}
N_{\mathrm{CE}} = \sum_{c \in \mathcal{C}} \sum_{j=1}^{7} n_{\mathrm{CE}_j}(c),
\qquad
f_{\mathrm{CE}_j} = \frac{\sum_{c \in \mathcal{C}} n_{\mathrm{CE}_j}(c)}{N_{\mathrm{CE}}}.
\label{eq:overall_compile_time_error_freq}
\end{equation}
This approach yields a single, combination-agnostic distribution for runtime and compile-time failures, without averaging per-row percentages.

\subsubsection{\textbf{Scenario and Pattern-Based Error Reduction Analysis (RQ3)}}
To understand where the algorithm helps, we compared the \emph{aggregate} frequency of each taxonomy subtype $t$ between Direct and Algorithm across all combinations (all models, datasets, and directions).

Let $\mathcal{C}$ be the set of combinations. Let $n_t^{\text{direct}}(c)$ and $n_t^{\text{algo}}(c)$ denote the \emph{raw counts} of failures labeled as type $t$ in combination $c$ under the direct and algorithm approaches, respectively. Then,
\begin{equation*}
\text{count}_{\text{direct}}(t) = \sum_{c \in \mathcal{C}} n_t^{\text{direct}}(c), \qquad
\text{count}_{\text{algo}}(t) = \sum_{c \in \mathcal{C}} n_t^{\text{algo}}(c),
\end{equation*}
\begin{equation*}
\text{count}_{\text{all}}(t) = \text{count}_{\text{direct}}(t) + \text{count}_{\text{algo}}(t).
\end{equation*}
The reported percentages are,
\begin{equation}
p_{\text{direct}}(t) = \frac{\text{count}_{\text{direct}}(t)}{\text{count}_{\text{all}}(t)}, \qquad
p_{\text{algo}}(t) = \frac{\text{count}_{\text{algo}}(t)}{\text{count}_{\text{all}}(t)}.
\label{eq:RQ3_mitigation_percentage}
\end{equation}
A taxonomy type is considered \emph{mitigated} when $p_{\text{Algo}}(t)$ is significantly lower than $p_{\text{Direct}}(t)$ (with the accompanying drop in raw frequency). This analysis was performed separately for compile-time and runtime subtypes.

\begin{figure*}
    \centering
    \includegraphics[width=7in]{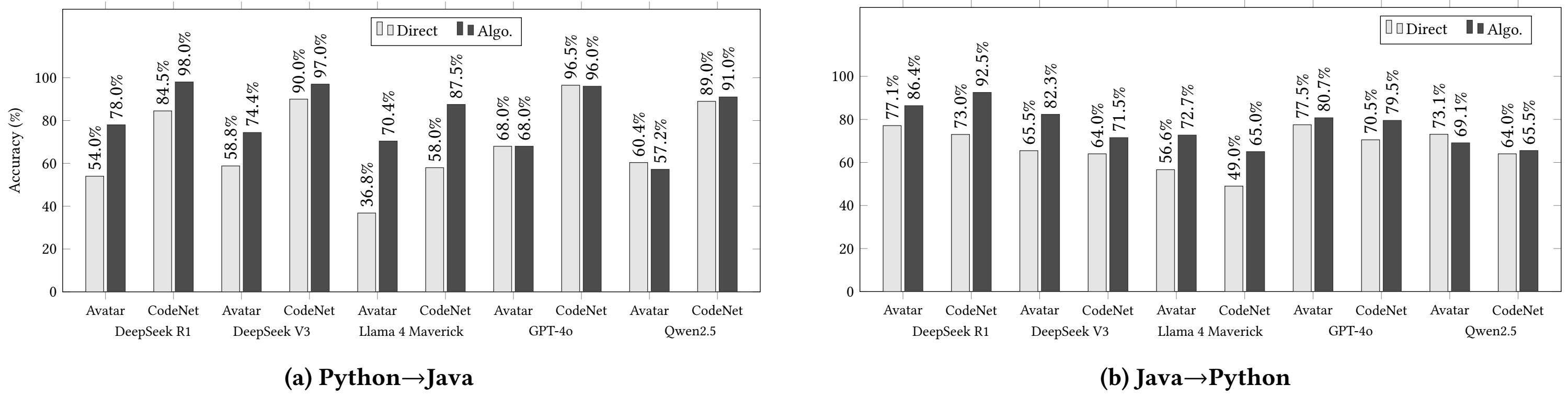}
    \caption{Accuracy comparison of Direct vs Algorithm-based (Algo.) approaches across models and datasets for both translation directions. Total per combination: Avatar (Python$\rightarrow$Java): $N{=}250$, Avatar (Java$\rightarrow$Python): $N{=}249$, CodeNet (Python$\rightarrow$Java, Java$\rightarrow$Python): $N{=}200$.}
    \label{fig:translation_accuracy_comparison}
    \vspace{-3mm}
\end{figure*}

\begin{table}[t]
\centering
\caption{Failure profile for Direct vs. Algo. across models, datasets, approach and translation directions. RE = Runtime error rate, CE = Compile-time error rate, TF = Test fail rate, TO = Timeout. Totals per combination: Avatar (Python$\rightarrow$Java): $N{=}250$, Avatar (Java$\rightarrow$Python): $N{=}249$, CodeNet (Python$\rightarrow$Java, Java$\rightarrow$Python): $N{=}200$. All values are in \%.}
\label{tab:performance_comparison_compact}
\setlength{\tabcolsep}{4pt}
\resizebox{\columnwidth}{!}{
\begin{tabular}{l | l | l | c c c c | c c c c}
\toprule
& & &
\multicolumn{4}{c|}{\textbf{Python$\rightarrow$Java}} &
\multicolumn{4}{c}{\textbf{Java$\rightarrow$Python}} \\
\cmidrule(lr){4-7}\cmidrule(lr){8-11}
\textbf{Model} & \textbf{Dataset} & \textbf{Appr.} &
\textbf{\makecell{RE}} &
\textbf{\makecell{CE}} &
\textbf{\makecell{TF}} &
\textbf{\makecell{TO}} &
\textbf{\makecell{RE}} &
\textbf{\makecell{CE}} &
\textbf{\makecell{TF}} &
\textbf{\makecell{TO}} \\
\midrule

\multirow{4}{*}{DeepSeek R1}
& \multirow{2}{*}{Avatar}  & Direct & 19.2 & 16.8 & 10 & 0 & 8.4 & 3.2 & 11.2 & 0 \\
&                           & Algo. & 8.8  & 0.8  & 12 & 0.4 & 8 & 0 & 5.6  & 0 \\
\cmidrule(lr){2-11}
& \multirow{2}{*}{CodeNet} & Direct & 6.5  & 9  & 0  & 0 & 10 & 5 & 12 & 0 \\
&                           & Algo. & 1  & 0.5  & 0.5  & 0 & 7  & 0 & 0.5 & 0 \\
\midrule

\multirow{4}{*}{DeepSeek V3}
& \multirow{2}{*}{Avatar}  & Direct & 8.8  & 14.4 & 17.2 & 0.8 & 24.9 & 2 & 7.6 & 0 \\
&                           & Algo.  & 10 & 0.8  & 14 & 0.8 & 12.1 & 0 & 4.8 & 0.8 \\
\cmidrule(lr){2-11}
& \multirow{2}{*}{CodeNet} & Direct & 2.5 & 6 & 1.5  & 0 & 34.5 & 1 & 0.5 & 0 \\
&                           & Algo. & 0.5 & 2 & 0.5  & 0 & 27.5 & 0.5 & 0 & 0.5 \\
\midrule

\multirow{4}{*}{Llama 4 Maverick}
& \multirow{2}{*}{Avatar}  & Direct & 50.4 & 1.2  & 11.6 & 0 & 18.5 & 6.4 & 18.5 & 0 \\
&                           & Algo. & 10.8 & 2.4  & 16 & 0.4 & 15.7 & 0.4 & 10 & 1.2 \\
\cmidrule(lr){2-11}
& \multirow{2}{*}{CodeNet} & Direct & 37 & 1.5  & 3.5  & 0 & 19.5 & 3.5 & 28 & 0 \\
&                           & Algo. & 4.5 & 1  & 7  & 0 & 26.5 & 1.5 & 6.5  & 0.5 \\
\midrule

\multirow{4}{*}{GPT-4o}
& \multirow{2}{*}{Avatar}  & Direct & 9.6 & 1.6 & 20.4 & 0.4 & 15.3 & 0 & 7.2 & 0 \\
&                           & Algo. & 10 & 2.4 & 19.2 & 0.4 & 12.1 & 0 & 7.2 & 0 \\
\cmidrule(lr){2-11}
& \multirow{2}{*}{CodeNet} & Direct & 0.5 & 1 & 2 & 0 & 28 & 0 & 1.5 & 0 \\
&                           & Algo. & 0.5 & 2.5 & 1 & 0 & 20 & 0 & 0.5 & 0 \\
\midrule

\multirow{4}{*}{Qwen2.5}
& \multirow{2}{*}{Avatar}  & Direct & 14.4 & 5.6 & 19.2 & 0.4 & 16.1 & 0 & 10 & 0.8 \\
&                           & Algo. & 14.4 & 7.6 & 20.4 & 0.4 & 18.1 & 2.4 & 8.8 & 1.6 \\
\cmidrule(lr){2-11}
& \multirow{2}{*}{CodeNet} & Direct & 2.5 & 4 & 4.5 & 0 & 29 & 1 & 5.5 & 0.5 \\
&                           & Algo. & 2 & 5.5 & 1.5 & 0 & 29 & 1.5 & 4 & 0 \\
\bottomrule
\end{tabular}
}
\vspace{-4mm}
\end{table}

\section{Study Findings}

In this section, we present the findings of our study based on both quantitative and qualitative analyses.

\subsection{Answering RQ1: Effective Performance of Algorithm-based Approach}

Figure \ref{fig:translation_accuracy_comparison} represents the accuracy comparison, and Table \ref{tab:performance_comparison_compact} presents the error profile for every model, dataset, and direction in both pipelines (direct vs. algorithm-based). We evaluated whether adding an explicit algorithm before generation improves translation quality. Overall, the algorithm-based workflow reduces compile-time and runtime failures, with the largest gains in cases that stress input/output (I/O) handling and typing.

The algorithm step lifts accuracy most where models benefit from committing to structure and I/O rules before emitting code. Llama 4 Maverick shows the largest gains on Python$\rightarrow$Java, moving from 36.8\% to 70.4\% on Avatar and from 58\% to 87.5\% on CodeNet, because the algorithm fixes type errors, loop bounds, and print formatting, which the direct translation often mishandles. DeepSeek R1 also improves strongly in all four settings, for example, from 54\% to 78\% on Avatar (Python$\rightarrow$Java) and from 73\% to 92.5\% on CodeNet (Java$\rightarrow$Python), reflecting its ability to turn a precise specification into consistent code. DeepSeek V3 improves across every combination, but with narrower margins, suggesting a solid direct baseline that still benefits from explicit contracts. GPT-4o changes little on Python$\rightarrow$Java because its direct baseline is already high, yet it gains on Java$\rightarrow$Python where an algorithm helps preserve parsing discipline and loop bounds. When the algorithm is concrete, Qwen2.5 helps by abstracting away token cues like identifiers or specific I/O idioms; accuracy can drop. Overall, the algorithm pays off for typing and I/O-Intensive tasks, letting the model lock in data shapes, numeric rules, and output formats.

The same pattern appears in the error breakdown. The algorithm reduces Java-side compile breaks by making setup explicit. DeepSeek R1 cuts compile errors on Python$\rightarrow$Java from 16.8\% to 0.8\%, and DeepSeek V3 similarly drops from 14.4\% to 0.8\%. Llama's big problem in Avatar (Python$\rightarrow$Java) is runtime instability, which the algorithm reduces from 50.4\% to 10.8\% by specifying input parsing and loop bounds. Across Java$\rightarrow$Python, compile errors often vanish for DeepSeek R1 and DeepSeek V3 once the algorithm removes type and entry-point assumptions. Runtime crashes fall when the algorithm names the input contract and numeric rules, for example GPT-4o on CodeNet drops from 28\% to 20\% and DeepSeek V3 on CodeNet from 34.5\% to 27.5\%. Residual failures cluster around parsing and conversion in Python targets, where tokenization and end-of-file handling are still easy to get wrong if the algorithm is not specific. Test-only mismatches usually decline when the algorithm fixes output ordering and separators, though a few trade-offs arise in which stricter compile-time or runtime checks surface new, smaller issues. Overall, the algorithm helps most by removing avoidable setup and I/O mistakes, leaving a smaller core of input parsing and coercion problems that future constraints and lightweight validators can address.

\textbf{Which approach is better:}  The algorithm-based pipeline is the reliable choice overall. It gives the biggest accuracy gains for Llama 4 Maverick and DeepSeek R1, steady gains for DeepSeek V3, and clear benefits for GPT-4o when the target is Python. Qwen2.5 is mixed and depends on the dataset. On errors, the algorithm most reliably avoids Java-side compile failures and often reduces runtime crashes when I/O and type errors matter. A few models drop slightly when the algorithm removes token-level cues they relied on. Overall, a brief language-agnostic algorithm before code translation improves structure and semantics, leading to fewer compile breaks, fewer runtime crashes, and fewer test-only mismatches.

\noindent \textbf{Overall micro-average accuracy.} We summarize each pipeline with a pooled metric following Eq.\ref{eq:micro_accuracy}:
For Direct, 
\[
\mathrm{Acc}_{\mathrm{micro}}(direct)=\frac{3043}{4495}=67.7\%,
\]
\indent For Algorithm,
\[
\mathrm{Acc}_{\mathrm{micro}}(algo)=\frac{3530}{4495}=78.5\%.
\]

\indent Thus, Algorithm improves the overall micro-average by
\[
\Delta_{\mathrm{micro}}=78.5\%-67.7\%=10.8\%
\]
These pooled results support the earlier findings: the algorithm-based approach is more balanced across models, datasets, and directions, with clearer gains when precise control flow, typing, and I/O contracts matter.

\textbf{Best-performing model:} DeepSeek R1 is the strongest overall with the algorithm-based pipeline. It reaches the highest accuracy in all four settings and shows a balanced error profile, with compile failures near zero, runtime failures down, and test failures low. It outperforms other models because it is a reinforcement-learned reasoning model, architecturally a high-capacity MoE, and empirically strong in code and math \cite{guo2025deepseek}, so it faithfully follows the algorithm and emits stable idiomatic target scaffolds. This combination of a clear algorithm and stable code generation explains its consistent wins across datasets and directions. 

\begin{table*}[t]
\centering
\caption{Error Frequency by Taxonomy
    \small{(RE1 = Dependency \& Entry-Point Issues,
    RE2 = Parsing \& Type Conversion Issues,
    RE3 = Index/Key Access Issues,
    RE4 = Missing State \& Invalid Reference, 
    RE5 = Arithmetic Errors, 
    RE6 = Resource Exhaustion, 
    CE1 = Import/Namespace Resolution,
    CE2 = Lexical \& Token Errors, 
    CE3 = Incomplete Constructs,
    CE4 = Structural \& Declaration Issues, 
    CE5 = Type/Overload Resolution Errors, 
    CE6 = Literal Constraints, 
    CE7 = Others)}.
}
\label{tab:error_frequency_by_taxonomy}
\resizebox{6in}{!}{
\begin{tabular}{l|l|l|p{3mm}p{3mm}p{3mm}p{3mm}p{3mm}p{4mm}|p{3mm}p{3mm}p{3mm}p{3mm}p{3mm}p{4mm}|p{3mm}p{3mm}p{3mm}p{3mm}p{3mm}p{3mm}p{4mm}|p{3mm}p{3mm}p{3mm}p{3mm}p{3mm}p{3mm}p{4mm}}
\toprule

& & &
\multicolumn{12}{c|}{\textbf{Runtime Error}} &
\multicolumn{14}{c}{\textbf{Compile-time Error}} \\
\cmidrule(lr){4-15} \cmidrule(lr){16-29}

\textbf{Model} & \textbf{Dataset} & \textbf{Appr.} &
\multicolumn{6}{c|}{\textbf{Python$\rightarrow$Java}} &
\multicolumn{6}{c|}{\textbf{Java$\rightarrow$Python}} &
\multicolumn{7}{c|}{\textbf{Python$\rightarrow$Java}} &
\multicolumn{7}{c}{\textbf{Java$\rightarrow$Python}} \\
\cmidrule(lr){4-9} \cmidrule(lr){10-15} \cmidrule(lr){16-22} \cmidrule(lr){23-29}
& & & RE1 & RE2 & RE3 & RE4 & RE5 & RE6 & RE1 & RE2 & RE3 & RE4 & RE5 & RE6 & CE1 & CE2 & CE3 & CE4 & CE5 & CE6 & CE7 & CE1 & CE2 & CE3 & CE4 & CE5 & CE6 & CE7\\
\midrule

\multirow{4}{*}{DeepSeek R1} 
& \multirow{2}{*}{Avatar} & Direct     & 35 & 6  & 3 & 1 & 2 & 1 & × & 14 & 4 & × & × & 3 & × & × & 4 & 38 & × & × & × &  × &  × & 1 & 7 & × & × & ×\\
&                        & Algo.  & 4  & 13 & 3 & 1 & 1 & × & × & 15 & 2 & × & × & 3 & × & × & × & 1 & × & × & 1 & × & × & × & × & × & × & ×\\
\cline{2-29}
& \multirow{2}{*}{CodeNet} & Direct    & 12 & × & × & 1 & × & × & × & 16 & 4 & × & × & × & × & × & × & 18 & × & × & × & × & 1 & 3 & 6 & × & × & ×\\
&                        & Algo.  & 1  & × & × & 1 & × & × & × & 13 & 1 & × & × & × & × & × & × & × & × & 1 & × & × & × & × & × & × & × & ×\\
\midrule

\multirow{4}{*}{DeepSeek V3} 
& \multirow{2}{*}{Avatar} & Direct     & × & 15 & 2 & 2 & × & 3 & × & 53 & 2 & 4 & × & 3 & 4 & 9 & 2 & 17 & 1 & 1 & 2 & × & × & 2 & 3 & × & × & ×\\
&                        & Algo.  & 3 & 15 & 3 & 4 & × & × & 4 & 20 & × & 4 & × & 2 & × & × & × & 1 & 1 & × & × & × & × & × & × & × & × & × \\
\cline{2-29}
& \multirow{2}{*}{CodeNet} & Direct    & 1 & 2 & × & 2 & × & × & 2 & 51 & 3 & 12 & × & 1 & 3 & 5 & × & 3 & × & 1 & × & × & × & × & 2 & × & × & ×\\
&                        & Algo.  & × & × & × & 1 & × & × & × & 49 & 3 & 3 & × & × & 3 & × & × & × & × & 1 & × & × & × & × & 1 & × & × & ×\\
\midrule

\multirow{4}{*}{Llama 4 Maverick} 
& \multirow{2}{*}{Avatar} & Direct     & 105 & 7 & 5 & 3 & 1 & 5 & 1 & 31 & 5 & 7 & × & 2 & 1 & × & × & × & 2 & × & × & × & × & × & 16 & × & × & ×\\
&                        & Algo.  & × & 12 & 7 & 6 & × & 2 & 3 & 27 & 6 & 1 & × & 2 & 3 & × & 1 & 1 & 1 & × & × & × & × & × & 1 & × & × & × \\
\cline{2-29}
& \multirow{2}{*}{CodeNet} & Direct    & 68 & × & 1 & 1 & × & 4 & × & 30 & 3 & 6 & × & × & × & × & 3 & × & × & × & × & × & × & 3 & 4 & × & × & ×\\
&                        & Algo.  & × & × & 1 & 7 & × & 1 & × & 38 & 4 & 10 & 1 & × & × & × & 1 & 1 & × & × & × & × & × & × & 3 & × & × & × \\
\midrule

\multirow{4}{*}{GPT-4o} 
& \multirow{2}{*}{Avatar} & Direct     & × & 15 & 5 & 1 & × & 3 & × & 27 & 3 & 5 & × & 3 & 2 & × & × & × & 1 & × & 1 & × & × & × & × & × & × & × \\
&                        & Algo.  & × & 15 & 4 & 3 & × & 3 & 1 & 23 & 1 & 2 & × & 3 & 1 & × & 1 & 3 & 1 & × & × & × & × & × & × & × & × & × \\
\cline{2-29}
& \multirow{2}{*}{CodeNet} & Direct    & × & × & × & 1 & × & × & × & 44 & 6 & 6 & × & × & × & × & × & × & 2 & × & × & × & × & × & × & × & × & ×\\
&                        & Algo.  & × & × & × & 1 & × & × & × & 28 & 4 & 8 & × & × & 3 & × & × & × & 2 & × & × & × & × & × & × & × & × & × \\
\midrule

\multirow{4}{*}{Qwen2.5} 
& \multirow{2}{*}{Avatar} & Direct     & 5 & 16 & 6 & 6 & × & 3 & × & 20 & 4 & 10 & 3 & 3 & 9 & × & × & × & 2 & 2 & 1 & × & × & × & × & × & × & × \\
&                        & Algo.  & 3 & 17 & 3 & 9 & × & 4 & 7 & 18 & 1 & 14 & × & 5 & 3 & × & × & 4 & 9 & × & 3 & × & × & × & 6 & × & × & × \\
\cline{2-29}
& \multirow{2}{*}{CodeNet} & Direct    & 2 & × & × & 3 & × & × & × & 44 & 4 & 9 & 1 & × & 2 & × & 1 & × & 1 & 1 & 3 & × & × & × & 2 & × & × & × \\
&                        & Algo.  & × & 1 & × & 2 & × & 1 & 2 & 41 & 1 & 13 & 1 & × & 2 & × & × & 3 & 3 & × & 3 & × & × & × & 3 & × & × & × \\
\bottomrule
\end{tabular}
}
\end{table*}

\begin{figure}
    \centering
    \includegraphics[width=3.4in]{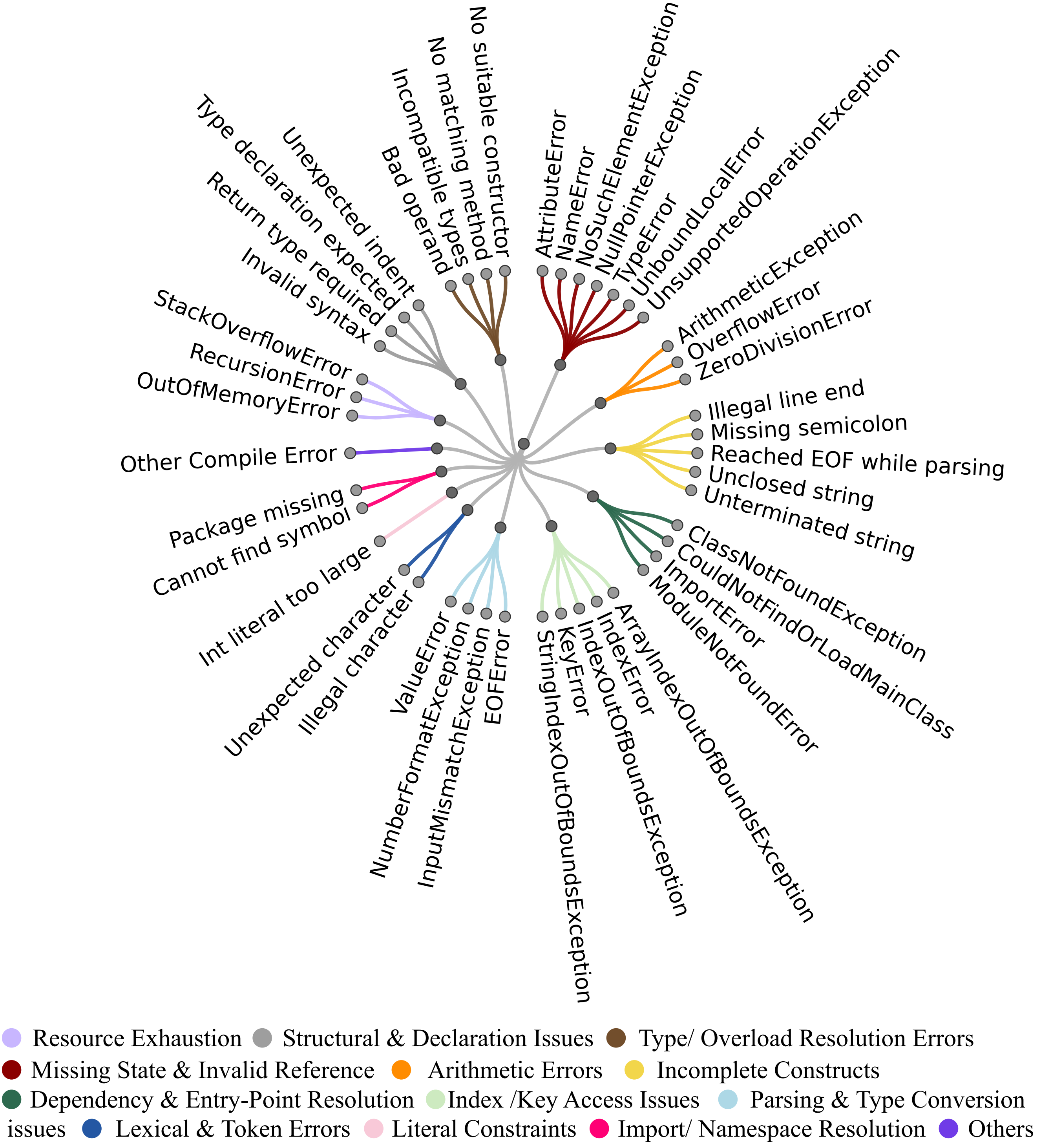}
    \caption{Error Taxonomy of Both Languages}
    \label{fig:taxonomies}
    \vspace{-4mm}
\end{figure}

\subsection{Answering RQ2: Error Taxonomy and Frequency Distribution Observed in Each Combination}

In our experiments, translated programs fail in four ways: runtime errors, compile-time errors, test failures, and timeouts. Timeouts are very rare, so we left them out. Test results show a clear pattern. For Python$\rightarrow$Java, direct and algorithm-based pipelines have similar test-fail rates. For Java$\rightarrow$Python, the algorithm-based pipeline fails fewer tests than the direct one. This suggests that moving from a statically typed language to a dynamically typed one benefits from an explicit algorithm step, which helps keep the intended behavior and output format. Since tests serve as external judges rather than diagnosing specific error types, they do not fit an error taxonomy. We analyzed compile-time and runtime failures, mapped them to language-aware subtypes, reported frequencies for both pipelines, and compared subtypes to see where the algorithm reduces errors.

\textbf{Constructing and analyzing the error taxonomy:} We grouped failures into two taxonomies: runtime and compile-time. Each group is a type of mistake, not a specific diagnosis. The definitions work for both Python$\rightarrow$Java and Java$\rightarrow$Python (Figure \ref{fig:taxonomies}).

For runtime failures, Dependency \& Entry-Point means the program cannot start or cannot find what it needs. Parsing \& Type Conversion covers input reading and value conversion when I/O styles or number rules differ. Index/Key Access covers out-of-range indexes and missing keys as bounds or map rules change. Missing State \& Invalid Reference means a value is unset, out of scope, not yet built, or the target blocks the operation. Arithmetic Errors covers illegal or unstable numeric work, including overflow and underflow. Resource Exhaustion covers unbounded recursion or growth that drains memory or the stack.

For compile-time failures, Import/Namespace Resolution indicates that packages or symbols are missing, with similar issues sometimes appearing later in dynamic targets. Lexical \& Token covers scanner errors such as illegal characters or unexpected tokens. Incomplete constructs mean unterminated or half-formed syntax when delimiters do not match. Structural \& Declaration Issues cover broken class, function, or module shapes, or missing required parts under stricter rules. Type/Overload Resolution means the compiler cannot match types or pick a method or a constructor. Literal Constraints covers invalid literal forms or ranges. Others is a small catch-all for all residual diagnostics.

These cause-based definitions support consistent aggregation (calculated by Eq. \ref{eq:overall_runtime_error_freq} and Eq. \ref{eq:overall_compile_time_error_freq}) across models, datasets, and directions (Table \ref{tab:error_frequency_by_taxonomy}) and set up the error reduction analysis that follows.

\textbf{Runtime failures:} For an overall view, we combined Avatar and CodeNet when counting failures. Across models, Parsing \& Type Conversion Issues (RE2) are the main runtime problem. DeepSeek R1 is typical here: RE2 is its largest bucket with 77 cases, well above any other category for that model. DeepSeek V3 shows the same pattern with 205 RE2 cases, which confirms that parsing inputs and coercing types are the hardest runtime steps. Llama 4 Maverick is the outlier. Its largest bucket is Dependency \& Entry-Point Issues (RE1) with 177 cases, which reflects frequent start-up and packaging mistakes. GPT-4o again aligns with the majority, with 152 in RE2, while Qwen2.5 is also parsing-led with 157 in RE2 and a secondary concentration in Missing State \& Invalid Reference (RE4) (66). Across all models, RE2 dominates with 736 cases, followed by RE1 (259) and RE4 (170). These patterns suggest that most models first struggle with reading and correctly converting inputs, while some additionally attempt to bootstrap the program and identify dependencies. 

\textbf{Compile-time failures:} When we combined datasets, Structural \& Declaration Issues (CE4) are the leading compile-time failure across models. DeepSeek R1 shows the clearest example with 70 CE4 cases, indicating frequent problems with required declarations or program structure in the target language. DeepSeek V3 also concentrates on CE4 (27) with a secondary presence of Lexical \& Token Errors (CE2). Llama 4 Maverick follows the same pattern with 26 in CE4 and far fewer elsewhere. GPT-4o is different: its compile breaks are mostly Import/Namespace Resolution (CE1) with 6 cases, consistent with a strong syntax baseline that still stumbles on package setup. Qwen2.5 spreads across CE4 (18), CE1 (16), and Type/Overload Resolution (CE5, 15), suggesting a mix of structure, imports, and typing issues. Aggregated across all models, CE4 is the clear headliner with 144 cases, followed by CE1 (36), CE5 (24), and CE3 (22). Overall, these results indicate that many compile failures come from not meeting the target language’s structural and declaration rules, while import resolution and typing are the next most common pain points.

Overall, the table \ref{tab:error_frequency_by_taxonomy} shows that runtime failures are dominated by Parsing \& Type Conversion (RE2, 55\%), which signals breakdowns in input handling and numeric coercion when language I/O idioms differ (for example, whitespace or tokenization rules, int or float parsing, integer division). The next most frequent is Dependency \& Entry-Point (RE1, 19.3\%), indicating programs that cannot start or locate required components after translation (for example, missing main, package, or module wiring). The remaining runtime groups appear less often: RE4 Missing State \& Invalid Reference 12.7\%, RE3 Index or Key Access 7.8\%, RE6 Resource Exhaustion 4.5\%, and RE5 Arithmetic 0.8\%. For compile-time failures, Structural \& Declaration Issues (CE4, 54.6\%) lead, reflecting malformed class or method scaffolding or missing declaration elements under stricter target rules (for example, return types, braces or semicolons, class headers). The second most common is Import or Namespace Resolution (CE1, 13.6\%), which points to unresolved packages or symbols during the build. Other compile groups occur at lower rates: CE5 Type or Overload Resolution 9.9\%, CE3 Incomplete Constructs 8.3\%, CE2 Lexical and Token Errors 5.7\%, CE7 Others 5.3\%, and CE6 Literal Constraints 2.7\%.

\begin{figure*}
    \centering
    \includegraphics[width=5in]{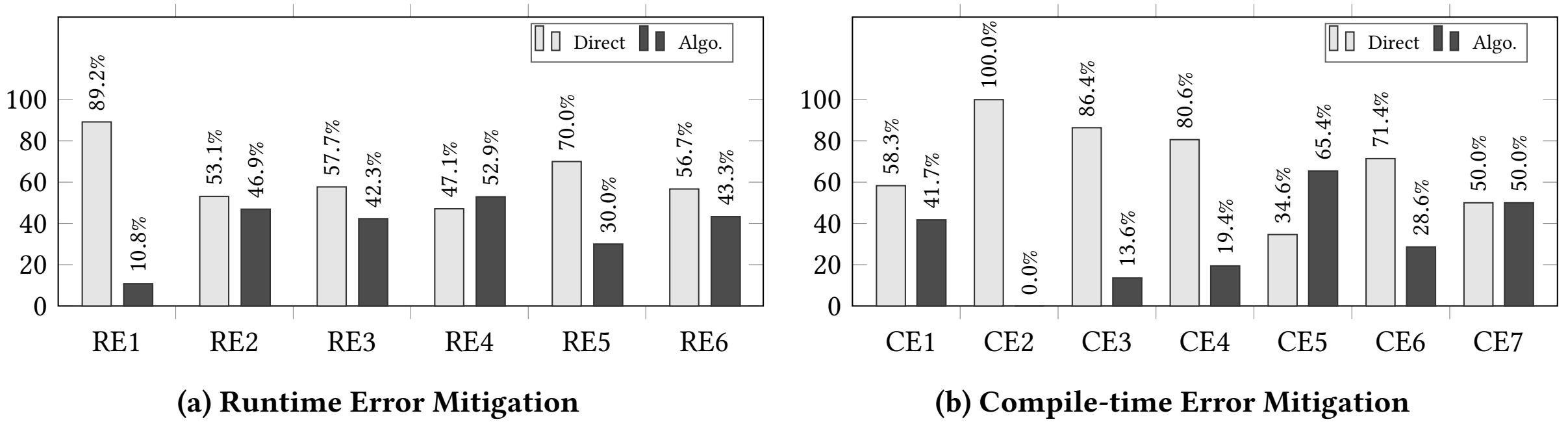}
    \caption{Algorithm-based vs Direct: per-type error mitigation
\small{
    (RE1 = Dependency \& Entry-Point Issues,
    RE2 = Parsing \& Type Conversion Issues,
    RE3 = Index/Key Access Issues,
    RE4 = Missing State \& Invalid Reference, 
    RE5 = Arithmetic Errors, 
    RE6 = Resource Exhaustion, 
    CE1 = Import/Namespace Resolution,
    CE2 = Lexical \& Token Errors, 
    CE3 = Incomplete Constructs,
    CE4 = Structural \& Declaration Issues, 
    CE5 = Type/Overload Resolution Errors, 
    CE6 = Literal Constraints, 
    CE7 = Others)}.
}

\label{fig:error-summary}
\vspace{-3mm}
\end{figure*}

\subsection{\textbf{RQ3: Scenario and Pattern-Specific Error Mitigation by Algorithm-based Approach}}
For the subsequent analysis, we computed the aggregate frequency of each taxonomy subtype using Eq.\ref{eq:RQ3_mitigation_percentage}.

The algorithm-based translation mitigates several runtime error categories relative to Direct (see Figure \ref{fig:error-summary}a). The clearest gain is in Dependency \& Entry-Point Issues, which fall from 231 of 259 (89.2\%) under Direct to 28 of 259 (10.8\%) under Algorithm, indicating that an explicit algorithm clarifies program startup configuration and required dependencies. Parsing \& Type Conversion Issues also improve from 391 of 736 (53.1\%) to 345 of 736 (46.9\%)-as the algorithm specifies input formats and value coercions that stabilize execution-time parsing. Index/Key Access Issues dropped from 60 of 104 (57.7\%) to 44 of 104 (42.3\%), consistent with the algorithm making bounds and access patterns explicit. Resource Exhaustion declines from 34 of 60 (56.7\%) to 26 of 60 (43.3\%), likely because the algorithm emphasizes termination conditions and growth limits. Arithmetic Errors decrease from 7 of 10 (70\%) to 3 of 10 (30\%) when numeric rules are stated up front. One category does not mitigate: Missing State \& Invalid Reference rises from 80 of 170 (47.1\%) to 90 of 170 (52.9\%), suggesting that the added structure introduced by the algorithm can expose initialization and scoping slips that were less frequent under Direct.

At compile-time, algorithm-based translation markedly reduces syntax- and structure-sensitive failures (Figure \ref{fig:error-summary}b). Lexical \& Token Errors collapse from 15 of 15 (100\%) under Direct to 0 of 15 (0\%) under Algorithm, showing that the algorithm enforces well-formed tokens. Incomplete constructs dropped from 19 of 22 (86.4\%) to 3 of 22 (13.6\%), reflecting clearer delimiter and block completion. Structural \& Declaration Issues shrink from 116 of 144 (80.6\%) to 28 of 144 (19.4\%) as the algorithm prescribes class/function skeletons and required signatures. Import/Namespace Resolution improves more modestly-from 21 of 36 (58.3\%) to 15 of 36 (41.7\%)-indicating partial standardization of imports with some environment-specific gaps remaining. Literal Constraints also improve, from 5 of 7 (71.4\%) to 2 of 7 (28.6\%), as the algorithm nudges literals toward target-legal forms. Two categories do not exhibit any reduction: Type/Overload Resolution Errors increase from 9 of 26 (34.6\%) to 17 of 26 (65.4\%), implying that the algorithm’s more explicit APIs surface mismatches earlier at compile-time, and Others remain unchanged at 7 of 14 (50\%) under both approaches.

\section{Implications}

\textbf{For practitioners and tool builders.}
Teams should adopt an algorithm-based workflow for translation rather than a one-shot approach. Empirically, this step improved overall accuracy by 10.8\% (67.7\% to 78.5\%), eliminated all lexical and token errors, and reduced incomplete constructs by 72.7\% and structural or declaration faults by 61.1\%. These gains translate to fewer compile breaks and runtime crashes that would otherwise require manual debugging. Developers should first prompt the model to produce a concise algorithm covering six items: (i) input handling, (ii) data structures, (iii) numeric rules, (iv) index bases and loop bounds, (v) termination, and (vi) output format. Treat this as a checklist--reject any plan that omits an item. Although generating this intermediate algorithmic plan adds one extra model call, results show it removes most preventable errors and lowers post-translation debugging and maintenance costs.

\textbf{For engineering practice at scale.}
Run translation as a deterministic pipeline rather than a single LLM call. Keep algorithm plans under version control, automatically validate them, and generate target code only after checks pass. In our study, this pipeline reduced dependency and entry-point failures by 78.4\% and cut runtime instability by over 50\%. Clear separation of compile-time, runtime, and test-failure stages accelerates regression diagnosis and makes model replacement safer, since validators remain stable even if backend models change. The small planning overhead pays back through lower defect density and faster recovery.

\textbf{For LLM and systems researchers.}
Remaining issues stem from weak enforcement of semantic constraints such as type resolution and initialization scope. Future work should make algorithms machine-checkable and integrate lightweight analyzers (type-width, indexing, and output-format checks). Controlled ablations in which constraints are toggled can reveal which constraints contribute most to reliability. Combining algorithm-guided generation with lightweight validation offers a practical path toward fully reliable, intent-preserving multilingual code translation.

\section{Threats to Validity}
Threat to \textit{external validity} concerns the generalizability of our results. We used Python and Java as representative high-contrast languages with dynamic and static typing to capture common translation challenges. Although the evaluation focused on this pair, the framework is language agnostic and can generalize to other language combinations with suitable prompt and test adaptations.

Threats to \textit{construct validity} concern whether our measures capture the intended constructs of translation failure. Error types were assigned manually, which introduced measurement subjectivity. We mitigated this with a written codebook that mentions inclusion and exclusion rules and direction-aware examples, a pilot double-coding pass, and precedence rules to avoid double-counting. Inter-rater agreement on the double-coded subset is Cohen's kappa = 0.893, and any remaining disagreements were resolved by discussion. These steps helped to ensure that the taxonomy is interpretable and applied consistently.

Threats to \textit{internal validity} concern whether observed differences actually arise from the algorithm step rather than from prompts, decoding settings, or randomness. We fixed prompts and decoding parameters for each condition, scripted the entire workflow, graded only the final code in both pipelines, standardized toolchains, and kept full logs. Some randomness can still occur. We reduced it by running evaluations within a short time window and by using fixed random seeds when the API allows.

The other \textit{internal validity} comes from the quality of the algorithm. It helped only when the algorithm is specific. Missing elements such as tokenization rules, end-of-file handling, data shapes, numeric policy, index bases, loop bounds, termination conditions, or output format can cause systematic errors. We mitigated this with a fixed algorithm in the prompt (see section \ref{section:algo_based_pipeline}) and by rejecting under-specified algorithm. This helped isolate the effect of planning from prompt design noise.

\section{Related Works}

\textbf{Traditional Code Translation.}
Early work in code translation relied on program analysis with hand-crafted transformation rules to migrate programs between languages. In general, the literature distinguishes two families of automated techniques: rule-based systems and data-driven (learning-based) methods. Representative rule-based tools include C2Rust \cite{immunant_c2rust} (translate code from C to Rust), CxGo \cite{gotranspile_cxgo} (translate code from C to Go), and Sharpen \cite{mono_sharpen} \& Java2CSharp \cite{paulirwin_JavaToCSharp_2025} (translate code from Java to CSharp). However, these systems are tightly coupled to each language. For every source-target pair, engineers must handwrite many rules for functions, objects, and standard libraries, making the approach difficult to scale \cite{yang2024exploring}. In addition, the resulting translations often lacks readability and may still exhibit correctness issues, further limiting their practical use \cite{nguyen2013lexical, nguyen2015divide, liu2023syntax}. Another line of work goes beyond hand-written rules and learns translation directly from data. These methods train on large code corpora, either supervised when parallel source-target pairs are available or unsupervised when only monolingual code is available. Over time, the field has shifted from statistical models \cite{nguyen2013lexical, nguyen2014migrating} to neural networks \cite{chen2018tree}, then to pre-trained representations \cite{wang2021codet5, zhu2024semi, szafraniec2022code, roziere2021leveraging, jiao2023evaluationofnmt} and now to large language models \cite{pan2024lost, yang2024exploring, macedo2025intertrans} reflecting a broader trend of using scale and learned abstractions to improve accuracy and coverage.

\textbf{Code Translation with LLMs.}
Instruction-tuned code LLMs trained on large, multilingual corpora now perform well on core software engineering tasks such as code translation, summarization, comprehension \cite{chen2021evaluating, nijkamp2022codegen, lozhkov2024starcoder, roziere2023code} and increasingly on program repair \cite{huang2023empirical, xia2023automated}, where they can locate faults, propose patches, and validate fixes with tests. Broad exposure to many languages helps these models internalize syntax and common semantic patterns, enabling transfer across tasks and targets. At the same time, strong prompt design and minimal adaptation enable general-purpose LLMs to compete closely on similar benchmarks \cite{Xiao2025LogicMM}. Although LLMs have improved rapidly, their translated code remains unreliable. Many open-source models score well on benchmarks but are not ready for production because they often miss the deeper semantics of a program \cite{pan2024lost}. Prior work tries to fix this after the fact, for example, by generating synthetic test cases \cite{yang2024exploring} or re-translating with compiler error messages \cite{pan2024lost, yang2024exploring}. These efforts help, but they do not directly test or strengthen the model's understanding of the intended computation \cite{pan2024lost, yang2024exploring}. This raises a central question: How much syntactic and semantic structure do these models preserve when translating code? Hence, unlike prior works that rely on post-hoc fixes or prompt tweaks, we intervened an explicit algorithm before generation to lock I/O contracts, data structures, control flow, and numeric rules, so the model grounds semantics before syntax.

\textbf{Intermediate Representations for Code Generation.}
A second line of work studies intermediate representations for code generation. Chain-of-Thought prompting elicits stepwise reasoning before emitting answers \cite{wei2022chain}, and Least-to-Most prompting decomposes hard tasks into simpler subproblems solved in sequence \cite{Zhou2022LeasttoMostPE}. Program-Aided Language Models (PAL) push this idea further by asking the model to produce executable programs as intermediate reasoning, offloading computation to a runtime \cite{gao2022pal}, while Program-of-Thoughts uses structured, program-like steps to improve reliability on symbolic tasks \cite{Chen2022ProgramOT}. Closer to our setting, a study proposes transitive intermediate translations, routing a source program through one or more existing programming languages before reaching the final target, and reports improved correctness when a direct translation struggles \cite{macedo2025intertrans}. In contrast to transitive multi-language routes or general planning, we used a single, language-agnostic algorithm tailored to source-to-source translation. To our knowledge, no prior approach systematically leveraged a language-agnostic "natural language algorithm", something LLMs are good at generating as an intermediate step to make the semantics of the code explicit before generating the actual code.

\section{Conclusion}
Code translation with LLMs has become increasingly practical, but direct one-shot generation still fails to preserve low-level program details essential to correctness. To address this, we introduced an \textit{algorithm-based pipeline} that inserts a concise, language-neutral specification step before code generation to capture I/O behavior, data structures, numeric rules, and loop bounds. We evaluated this approach across five LLMs on Avatar and CodeNet, translating both Python$\rightarrow$Java and Java$\rightarrow$Python, and automatically tested each translation for compilation, runtime stability, and correctness. Results show that the algorithmic step substantially improves reliability, raising the pooled micro-average accuracy from 67.7\% to 78.5\% and reducing major failure modes: compile-time Lexical and Token errors dropped to zero, Incomplete Constructs declined by 72.7\%, Structural and Declaration issues by 61.1\%, and runtime Dependency and Entry-Point failures by 78.4\%. DeepSeek R1 achieved the highest absolute gains, Llama 4 Maverick showed the largest relative improvement, GPT-4o benefited most on Java$\rightarrow$Python, and Qwen2.5 revealed where algorithms must be more specific. Overall, these findings establish the algorithm-based pipeline as a practical and general principle for reliable, intent-preserving code translation with LLMs. Although the extra planning step introduces minor computational overhead, it eliminates many preventable errors and reduces long-term debugging and maintenance costs. Future work will extend this framework to additional language pairs and multi-file projects, integrate property-based and semantic testing, and develop machine-checkable plan schemas for automated validation.

\begin{acks}
This research is supported in part by the Natural Sciences and Engineering Research Council of Canada (NSERC) Discovery Grants program, the Canada Foundation for Innovation's John R. Evans Leaders Fund (CFI-JELF), and by the industry-stream NSERC CREATE in Software Analytics Research (SOAR).
\end{acks}

\balance
\bibliographystyle{ACM-Reference-Format}
\bibliography{references}

\end{document}